\documentstyle[12pt]{article}

\newtheorem{definition}{Definition}
\newtheorem{thm-def}{Definition-Theorem}
\newtheorem{thm}{Theorem}
\newcommand{\s}{\sigma}
\renewcommand{\t}{\tau}
\newcommand{\p}{product integral}

\newcommand{\eq}{equation}
\newcommand{\beq}{\begin{equation}}
\newcommand{\eeq}{\end{equation}}

\title{\nopagebreak
\begin{flushright}
\tenrm UCTP101.99
\end{flushright}\vskip0.3in
\nopagebreak
\large \bf  Product Integral Representations of \\
Wilson Lines and Wilson loops \\
and Non-Abelian Stokes Theorem}
\author{Robert L. Karp\thanks{e-mail address:
karp@physics.uc.edu},\,
Freydoon Mansouri\thanks{e-mail address: mansouri@uc.edu}\\
{\it \small
Physics Department, University of
Cincinnati, Cincinnati, OH 45221}\\
Jung S. Rno\thanks{e-mail address: rno@uc.edu} \\
{\it \small
Physics Department, University of Cincinnati-RWC, Cincinnati, OH
45236}}
\date{}

\begin{document}
\maketitle

\begin{abstract}
We make use of product integrals to provide an unambiguous
mathematical representation of Wilson line and Wilson loop
operators. Then, drawing upon various properties of product
integrals, we discuss such properties of these operators
as approximating them with partial sums, their convergence,
and their behavior under gauge transformations.
We also obtain a surface product integral representation for the
Wilson loop operator. The result can be interpreted as the
non-abelian
version of Stokes theorem.
\end{abstract}

\section{Introduction}

The notion of Wilson loop~\cite{rone,rtwo} provides a systematic
method of
obtaining gauge invariant observables in gauge theories. Its
applications
range over such diverse fields as phenomenology and lattice gauge
theories
on the practical side and topological gauge
theories~\cite{rthree} and
string theory~\cite{polch} on the purely theoretical side. The
importance
of the Wilson line as a parallel transport operator in the gauge
independent formulation of gauge theories has been emphasized by
Mandelstam~\cite{mandl}, and further developed by Wu and
Yang~\cite{yang}.
More recently, in the context of the AdS/CFT
correspondence~\cite{mal1}, an
interesting connection between Wilson loops in supersymmetric
gauge
theories and membranes in supergravity theories has been
suggested~\cite{mal}. In view of all these developments, it is
imperative
that Wilson lines and Wilson loops be described within a well
defined
mathematical framework. The main purpose of this work is to
provide such a
representation by means of product integrals. This will permit us
to give,
among other things, two unambiguous proofs of the non-abelian
version of
Stokes theorem.

The product integral formalism has been used extensively in the
theory of
differential equations and of matrix valued
functions~\cite{rfour}. In the
latter context, it has a built-in feature for keeping track of
the {\it
order} of the matrix valued functions involved. As a result,
product
integrals are ideally suited for the description of {\it path
ordered}
quantities such as Wilson lines and Wilson loops. In fact, they
make
precise what one means by these concepts as well as what one
means by the
notion of path ordering in general.  Moreover, since the theory
of product
integrals is well developed independently of particular
applications, we
can be confident that the properties of Wilson lines and Wilson
loops
which we establish using this method are correct and unambiguous.

Among the important advantages of the product integral
representation of
Wilson lines, one is the manner in which it deals with
convergence issues.
In the physics literature, the exponential of an operator such as
a Wilson
line is defined formally in terms of its power series expansion.
In such a
representation, it will be difficult, without any further
elaboration, to
establish whether the series converges and if so how. In contrast
to this,
it is a straight forward matter to establish the criteria for the
convergence of the Wilson line in its product integral
representation. This is because in such a framework
the Banach space structure of the corresponding matrix valued
functions is already built into the formalism.

Another important advantage of the product integral formulation
of the
Wilson loop is that, at least for orientable surfaces, it permits
a
2-surface representation for it. Based on the central role of
Stokes
theorem in physics and in mathematics, it is not surprising that
the
non-abelian version of this theorem has attracted a good deal of
attention
in the physics literature~\cite{rfive}-~\cite{r15}. The central
features
of the earlier attempts\cite{rfive}-~\cite{r11} have been
reviewed and
improved upon in a recent work~\cite{r12}. Other recent works on
non-abelian Stokes theorem~\cite{r13,r14,r15} focus on specific
problems
such as confinement~\cite{r14}, zig-zag symmetry~\cite{r15}
suggested
Polyakov~\cite{r16}, etc. With one exception~\cite{r12}, the
authors of
these works seem to have been unaware of a 1927 work in the
mathematical
literature by Schlessinger~\cite{r17} which bear strongly on the
content
of this theorem. Modern non-abelian gauge theories did not exist
at the
time, and Schlesinger's work dealt with integrals of matrix
valued
functions and their ordering problems. Its relevance to Wilson
lines and
Wilson loops is tied to the fact that in non-abelian gauge
theories, the
connection and curvature are matrix valued functions. As a
result,
Schlesinger's work amounts to establishing the non-abelian Stokes
theorem
in two (target space) dimensions. By an appropriate extension and
reinterpretation of his results, we show that the product
integral
approach to the proof of this theorem is valid in any target
space
dimension.

This work is organized as follows: To make this manuscript self-
contained, we review in Section 2 the main
features of product integration~\cite{rfour} and state without
proof a
number of theorems which will be used in the proof of the
non-abelian
Stokes theorem and other properties. In Section 3, we express
Wilson lines and Wilson loops in terms of product integrals.
In Section 4, we turn to the
proof of the non-abelian Stokes
theorem for
orientable surfaces. In section 5, we give a variant of this
proof.
Section 6 is devoted to convergence issues for Wilson lines and
Wilson
loops. Section 7 deals with two observables from the Wilson loop
operator.
In Section
8, we study behavior of Wilson lines and Wilson loops under gauge
transformations. This is another instance in which the
significant
advantage of the product integral representation of these
operators
becomes transparent.

\section{Some properties of product integrals}

The method of {\em product integration} has a long history, and
its origin
can be traced to the works of Volterra\cite{rfour}.  The
justification for
its name lies in the property that the \p\ is to the product what
the
ordinary (additive) integral is to the sum. One of the most
common
applications of product integrals is to the solution of systems
of linear
differential equations. To see how this comes about, let us
consider an
evolution equation of the type
\beq
{\bf Y'(s)} = A(s){\bf Y(x)},\quad {\bf Y(s_0)} = {\bf Y_0}.
\label{e1}
\eeq
where s is a real parameter, and prime indicates differentiation.
When the quantities $Y$ and $A$ are
ordinary functions, and $Y_0$ is an ordinary number, the solution
is given
by an ordinary integral. On the other hand, if these quantities
are matrix
valued functions arising from a system of, say, $n$ linear
differential
equations in $n$ unknowns, then the solution will be a product
integral.

To motivate a more precise formulation of product integration, we
start with a simple example which exhibits its main features.
Let us suppose that all the matrix valued
functions that appear in the above \eq\ are continuous on the
real
interval $[a,b]$.  Then, given the value of ${\bf Y}$ at the
point $a$,
i.e., given ${\bf Y}(a)$, we want to find ${\bf Y}(b)$. One can
obtain an
approximate value for ${\bf Y}(b)$ using a variant of Euler's
tangent-line
method. Let $P=\{s_0,s_1,\ldots,s_n\}$ be a partition of the
interval
$[a,b]$, and let $\Delta s_k=s_k-s_{k-1}$ for all $k=1,\ldots,n$.
In the
interval $[s_0,s_1]$, we approximate $A(s)$ by the constant value
$A(s_1)$, solve the differential equation with initial value
${\bf Y}(a)$
and get the approximate solution for ${\bf Y}$ at $ s_1$:
$$
{\bf Y}(s_1)\approx e^{A(s_1)\Delta s_1}{\bf Y}(a).
$$

In the next interval $[s_1,s_2]$, using the above approximate
value as
input, and replacing $A(s)$ by $A(s_2)$, one finds
$$
{\bf
Y}(s_2)\approx e^{A(s_2)\Delta s_2}e^{A(s_1)\Delta s_1}{\bf
Y}(a).
$$
Proceeding in this manner we obtain the following approximate
value
for
${\bf Y}(b)$:
\beq
{\bf Y}(b)\approx e^{A(s_n)\Delta s_n}\ldots e^{A(s_1)\Delta
s_1}{\bf Y}(a) = \Pi_p(A) {\bf Y}(a),
\label{e2}
\eeq
where
$$
\Pi_p(A)=\prod_{k=1}^{n}e^{A(s_k)\Delta s_k}.
$$
We stress that the order of the exponentials on the right hand
side of
this equation is important since the corresponding matrices do
not commute
in general.

Since $A$ is continuous on the compact interval $[a,b]$, it
follows that
$A$ is uniformly continuous in that interval. This implies that
for all
$k=1,\ldots,n$ the value $A(s_k)$ will be close to the values of
$A(s)$ on
$[s_{k-1},s_k]$. It is thus reasonable to suppose that if the
mesh
$\mu(P)$ of the partition $P$ (the length of the longest
subinterval) is
small, the above calculation results in a good approximation to
${\bf
Y}(b)$. Then, we expect that the exact value of ${\bf Y}(b)$ is
given by
the natural limiting procedure
\beq
{\bf Y}(b)=\lim_{\mu(P)\rightarrow 0} \Pi_P(A) {\bf Y}(a)
\equiv \prod_{a}^{x} e^{ A(s)ds} {\bf Y}(a).
\eeq

Having identified the main ingredients which characterize the
above
construction, we proceed to give a precise definition of the
product integral~\cite{rfour}. We begin
with the

\begin{definition}

Let $A:[a,b]\rightarrow {\bf C}_{n\times n}$ be a function with
values in
the space of complex $n\times n$ matrices.  Let
$P=\{s_0,s_1,\ldots,s_n\}$
be a partition of the interval $[a,b]$, with $\Delta
s_k=s_k-s_{k-1}$ for
all $k=1,\ldots,n$.
\begin{description}
\item[\it (i)] A is called a {\em step function} iff there is a
partition
$P$ such that A is constant on each open subinterval
$(s_{k-1},s_k)$ for
all $k=1,\ldots,n$.

\item[\it (ii)]The {\em point value approximant } $A_P$
corresponding to
the function A and partition P is the step function taking the
value
$A(s_k)$ on the interval $(s_{k-1},s_k]$ for all $k=1,\ldots,n$.

\item[\it (iii)]If $A$ is a step function, then we define the
function
$E_A:[a,b]\rightarrow {\bf C}_{n\times n}$ by $E_A(x):=
e^{A(s_k)(x-s_{k-1})} \ldots e^{A(s_2)\Delta s_2} e^{A(s_1)\Delta
s_1}$
for any $x\in (s_{k-1},s_k]$, for all $k=1,\ldots,n$ and
$E_A(a):=I$.
\end{description}
\end{definition}
After a number of intermediate developments, one arrives at the
following
fundamental theorem which can be taken as the starting point of
product
integration:

\begin{thm-def}

Given a continuous function $A:[a,b]\rightarrow {\bf C}_{n\times
n}$ and a sequence of step functions
$\{A_n\}$, which converge to $A$ in the sense of 
$L^1([a,b])$, then the sequence $\{E_{A_{n}}(x)\}$
converges
uniformly on $[a,b]$ to a matrix called the {\em \p }\ of A over
$[a,b]$.

\end{thm-def}
More explicitly, we have:
\beq
The\;\; product\;\; integral\;\; of\;\; A\;\; over \;\;
[a,b]\quad =\quad
\prod_{a}^{b} e^{ A(s)ds}.
\eeq

Now we are in position to enumerate some of the basic properties
of
product integrals. The proofs of these assertions are given in
reference~\cite{rfour}. Let $A:[a,b]\rightarrow {\bf C}_{n\times
n}$ be a
continuous function, and for any $x\in [a,b]$ let
\beq
F(x,a):=\prod_{a}^{x} e^{ A(s)ds}
\eeq
denote the \p\ from $a$ to $x$.  Then, $F$ satisfies the
following integral equation:
\beq
F(x,a)=1+\int_{a}^{x}\,ds\,A(s)F(s,a),
\label{ei}
\eeq
where $I=I_{n\times n}$ is the $n\times n$ unit matrix.
The quantity $F$ is also a solution of the following initial
value problem:
\beq
\frac{d}{dx}F(x,a)=A(x)F(x,a),\quad F(a,a)=I.
\label{eii}
\eeq
Although product integrals can formally be defined for singular
matrices,
the above definition makes sense if they are non-singular. This
is true,
e.g., when the matrices form a group.  Then the determinant of
the product
integral is given by the following theorem:

\begin{thm}

Given the continuous function $A:[a,b]\rightarrow {\bf
C}_{n\times n}$,
then for every $x\in [a,b]$, the \p\ $\prod_{a}^{x} e^{ A(s)ds}$
is non-singular and the following formula holds:
\beq
\det\left(\prod_{a}^{x} e^{ A(s)ds}\right)=e^{ \int_{a}^{x}\, tr
A(s)ds},
\eeq
\label{thm1}
where ``tr'' stands for trace.
\end{thm}

When the set of matrices $\{ A(s):s\in [a,b]\}$ is commutative,
i.e.
$[A(s),A(s')]=0\quad \forall s,s' \in [a,b]$, it is easy to show
that
\beq
\prod_{a}^{x} e^{ A(s)ds}=e^{ \int_{a}^{x}\, A(s)ds} .
\eeq

It is convenient to define the \p\ $\prod_{a}^{b} e^{ A(s)ds}$
also in the
case when $a\geq b$. It will be recalled that for ordinary
(additive) integrals
$\int_{a}^{b}\, A(s)ds=-\int_{b}^{a}\, A(s)ds$. To obtain the
analog of this for product integrals, we merely
replace the "additive" property with the corresponding
"multiplicative" property:
\beq
\prod_{a}^{b} e^{ A(s)ds}:= \left( \prod_{b}^{a} e^{
A(s)ds}\right)^{-1}.
\eeq
The additive property of ordinary integrals also provides a
composition rule for them:
 $\int_{a}^{c}\, A(s)ds+\int_{c}^{b}\, A(s)ds=\int_{a}^{b}\,
A(s)ds.$ For product integrals, we have an analogous composition
rule~\cite{rfour}:
\beq
\prod_{a}^{b} e^{ A(s)ds}=\prod_{c}^{b} e^{ A(s)ds}\prod_{a}^{c}
e^{A(s)ds}.
\eeq
Another well known property of ordinary integrals is the
differentiation rule with
respect to the endpoints: ${\partial\over{\partial b}}\left(
\int_{a}^{b}
\, A(s)ds \right) =A(b)$ and ${\partial\over{\partial a}}\left(
\int_{a}^{b} \, A(s)ds\right)=-A(a)$. The following theorem gives
the corresponding rule for \p s.

\begin{thm}
\label{thm2}

Let $A:[a,b]\rightarrow {\bf C}_{n\times n}$ be continuous. For
any
$x,y\in [a,b]$ we have:
\beq
\frac{\partial}{\partial x}\left(\prod_{y}^{x} e^{ A(s)ds}\right)
=A(x)\prod_{y}^{x} e^{ A(s)ds},\quad
\frac{\partial}{\partial y}\left(\prod_{y}^{x} e^{ A(s)ds}\right)
=
-\prod_{y}^{x} e^{
A(s)ds} A(y).
\eeq
\end{thm}
It is important to keep in mind the relative order of the $A(x)$
and $A(y)$
with respect to the \p.

The usual elementary way of computing ordinary integrals is by
means of the fundamental theorem calculus: $\int_{a}^{b} \,
f(x)ds=F(b)-F(a)$, where $F$
is a primitive function of $f$ ($F'=f$). To obtain the
corresponding theorem for product integrals, we start
by defining the so called L-operation which is a generalization
of the
logarithmic derivative, for non-singular functions:
\begin{definition}
The L-derivative of a non-singular differentiable function
$P:[a,b]\rightarrow {\bf C}_{n\times n}$ is given by:
\beq
LP(x):=P'(x)P^{-1}(x).
\eeq
\end{definition}
To demonstrate the usefulness of this operation, let us consider
the \p\
$P(x)=\prod_{a}^{x} e^{ A(s)ds}P(a)$.  From Theorem \ref{thm2},
we have
$P'(x)=A(x)\prod_{a}^{x} e^{ A(s)ds}P(a)=A(x) P(x)$. Then, from
the above
definition, we get $(LP)(x)=A(x)$ (the derivative of the
primitive
function is the
original function). We are thus led to the analog of the
fundamental
theorem of calculus for product integrals~\cite{rfour}:
\begin{thm}
\label{thm3}
For a non-singular and continuously differentiable ($C^1[a,b]$)
function $P:[a,b]\rightarrow {\bf C}_{n\times n}$, we
have
\beq
\prod_{a}^{x} e^{ (LP)(s)ds}=P(x)P^{-1}(a).
\eeq
\end{thm}
The following elementary properties of the $L$-operation follow
from its definition:
\beq
(LP^{-1})(x)=(P^{-1})'(P^{-1})^{-1}=(-P^{-1}P'P^{-1})P=-P^{-1}(x)
P'(x),
\eeq
and
\beq
L(PQ)(x)=(P'Q+PQ')Q^{-1}P^{-1}=LP(x)+P(x)(LQ(x))P^{-1}(x).
\eeq
We will rely heavily on the contents of the next three theorems
in proving the non-abelian version of Stokes theorem. The proofs
are
given in reference~\cite{rfour}.

\begin{thm}
\label{thm4}

({\em Sum rule}): Given continuous functions
$A,B:[a,b]\rightarrow {\bf C}_{n\times n}$,
let $ P(x)=\prod_{a}^{x} e^{ A(s)ds}$. Then
\beq
\prod_{a}^{x} e^{\, [A(s)+B(s)]ds}=P(x)\prod_{a}^{x} e^{
\,P^{-1}(s)B(s)P(s)ds}.
\eeq
\end{thm}

\begin{thm}
\label{thm5}

({\em Similarity rule}): Given a continuous function
$B:[a,b]\rightarrow {\bf C}_{n\times
n}$ and the non-singular function $P:[a,b]\rightarrow {\bf
C}_{n\times n}$, then
\beq
P(x)\left( \prod_{a}^{x} e^{
B(s)ds}\right)P^{-1}(a)=\prod_{a}^{x}
e^{\, [LP(s)+P(s)B(s)P^{-1}(s)]ds}.
\eeq
\end{thm}

\begin{thm}

({\em Derivative with respect to a parameter}): Given a function
$A:[a,b]\times[c,d]\rightarrow {\bf C}_{n\times n}$ such
that $A(s,\lambda)$ is continuous in $s$ for each fixed
$\lambda\in[c,d]$.
and is differentiable with respect to $\lambda$.
Then, the \p $P(x,y;\lambda)=\prod_{y}^{x} e^{
A(s;\lambda)ds}$ is differentiable with respect to $\lambda$, and
\beq
\frac{\partial}{\partial \lambda} P(x,y;\lambda)=\int_{y}^{x}
d s\,P(x,s;\lambda)\frac{\partial A}{\partial
\lambda}(s;\lambda)P(s,y;\lambda).
\eeq
\label{thmp}
\end{thm}
To put the above description in its proper perspective, we note
that instead of the specific complex Banach space $L^1 ([a,b ]$,
\p s can be defined over
more general Banach spaces. Consider, e.g., a set ${\cal B}(X)$
of bounded linear operators over 
a complex Banach space,
and let $A:[a,b]\rightarrow {\cal B}(X)$ be an operator valued
function. It is
possible to define the \p\ of $A$ and establish the analogs of
the properties given above in this more general context.
Then, the standard topologies (norm, strong,
and weak) on the space of bounded linear operators play an
important 
role. Moreover, the notion of Lebesgue integrable functions used
on 
$L^1([a,b])$ space above generalize naturally to
Boschner integrable functions \cite{bosch}. For details we refer
again
to \cite{rfour}.

\section{The representation of Wilson Line and Wilson Loop}

As noted in the introduction, Wilson lines and Wilson loops are
intimately related
to the
structure of non-abelian gauge theories.  To provide the
background for
using the product integral formalism of Section 2 to explore
their
physical properties, we begin with the statement of the problem
as it
arises in the physics context. Let $M$ be an n-dimensional
manifold
representing the space-time (target space). Let $A$ be a
(connection)
1-form on $M$. When $M$ is a differentiable manifold, we can
choose a
local basis $dx^{\mu}$, $\mu=1,...,n$, and express $A$ in terms
of its
components:
$$
A(x) = A_{\mu}(x) \; dx^{\mu}.
$$
We take $A$ to have values in the Lie-algebra, or a
representation thereof,
of a Lie group. Then, with $T_k$, $k=1,..,m$, representing the
generators
of the Lie group, the components of $A$ can be
written as
$$
A_{\mu}(x) = A_{\mu}^k (x)\; T_k.
$$
With these preliminaries, we can express
the  Wilson line of the non-abelian gauge theories in the
form~\cite{r18}
$$
W_{ab}(C) = {\cal P} e^{\int_{a}^{b} A}.
$$
where ${\cal P}$ indicates path ordering, and $C$ is a path in
$M$. When the path $C$ is closed, the corresponding Wilson line
becomes a Wilson loop~\cite{r18}:
\beq
W(C) = {\cal P} e^{\oint A}.
\eeq
The path $C$ in $M$ can be described in terms of an intrinsic
parameter $\s$, so that for points of $M$ which lie on the
path $C$, $x^{\mu} = x^{\mu}(\s)$. One can then write
$$
A_{\mu}(x(\s))dx^{\mu} = A(\s)d\s ,
$$
where
$$
A(\s) \equiv A^{\mu}(x(\s))\frac{dx^{\mu}(\s)}{d\s}.
$$
It is the quantity $A(\s)$, and the variations thereof, which we
will
identify with the matrix
valued functions of the product integral formalism.

Let us next consider the Wilson loop. For simplicity, we
assume that
$M$ has trivial first homology group with integer coefficients,
i.e.,
$H_1(M,{\bf Z})=0$. This
insures that the
loop may be taken to be the boundary of a two dimensional surface
$\Sigma$
in $M$. More explicitly, we take the 2-surface to be an
orientable
submanifold of $M$. It will be convenient to describe the
properties of
the 2-surface in terms of its intrinsic parameters $\s$ and $\t$
or $\s^a$,
$a=0,1$. So, for the points of the manifold $M$, which lie on
$\Sigma$, we
have $x = x(\s, \t)$. The components of the 1-form $A$ on
$\Sigma$ can be
obtained by means of the vielbeins (by the standard pull-back
construction):
$$
\it{v^{\mu}_a} = \partial_a \; x^{\mu}(\s).
$$
Thus, we get
$$
A_a = \it{v^{\mu}_a} \; A_{\mu}.
$$
The curvature 2-form $F$ of the connection $A$ is given by
$$
F = dA + A \wedge A ={1\over 2} F_{\mu \nu} \; dx^{\mu}\wedge
dx^{\nu}.
$$
The components of $F$ on $\Sigma$ can again be obtained by means
of the
vielbeins:
$$
F_{ab} = \it{v^{\mu}_a} \; \it{v^{\nu}_b}\;F_{\mu \nu}.
$$
We note at this point that we can construct the pulled-back
field
strength $F_{ab}$  in another way, as the the field strength of
the
pulled-back
connection $A_a$. It is easy to check that these two results
coincide,
insuring the consistency of the construction.

We want to express the Wilson loop operator in terms of product
integrals~\cite{r20}. To achieve this,
we begin with the
definition of a Wilson line in terms of a product integral.
Consider the continuous map
$A:[s_0,s_1]\rightarrow {\bf C}_{n\times n}$
where $[s_0,s_1]$ is a real interval. Then, we define the
Wilson line given above in terms of a product integral as
follows:
$$
{\cal P} e^{\int_{s_0}^{s_1} A(s)ds}
\equiv \prod_{s_0}^{s_1} e^{A(s)ds}.
$$
Anticipating that we will identify the closed path $C$ over which
the Wilson loop is defined with the boundary of a 2-surface, it
is convenient to work from the beginning with the matrix valued
functions $A(\s, \t)$. This means that our
expression for the Wilson line will depend on a parameter. That
is, let
\beq
A:[\s_0,\s_1]\times[\t_0,\t_1]\rightarrow {\bf C}_{n\times n},
\eeq
where $[\s_0,\s_1]$ and $[\t_0,\t_1]$ are real intervals on the
two surface $\Sigma$ and hence in $M$.
Then, we define a
Wilson line
\beq
P(\sigma,\s_0;\t)= \prod_{\s_0}^{\s} e^{
A_1(\sigma';\t)d\sigma'}
\equiv {\cal P} e^{\int_{\s_0}^{\s} A_1(\sigma';\t)d\sigma'}.
\label{dp}
\eeq
In this expression, ${\cal P}$ indicates path ordering with
respect to $\s$, while $\t$ is a parameter. To be able to
describe
a Wilson loop, we similarly define the Wilson line
\beq
Q(\sigma;\t,\t_0) = \prod_{\t_0}^{\t} e^{ A_0(\sigma;\t')d\t'}
\equiv {\cal P} e^{\int_{\t_0}^{\t} A_0(\s;\t')d\t'}.
\label{dq}
\eeq
In this case, the path ordering is with respect to $\t$, and $\s$
is a parameter.

To prove the non-abelian version of the Stokes theorem, we
want to make use of product integration techniques to express the
Wilson loop operator as an integral over a two dimensional
surface bounded by the corresponding loop. In terms of the
intrinsic coordinates of such a surface, we can write the Wilson
loop operator in the form
\beq
W(C) = {\cal P} e^{\oint A_a d\s^a},
\label{s24}
\eeq
where, as mentioned above,
\beq
\s^a=(\t,\s)\, ; \quad a=(0,1).
\eeq
The expression for the Wilson loop depends on the homotopy class
of
paths in $M$ to
which the closed path $C$ belongs. We can, therefore,
parameterize the path $C$ in any convenient manner consistent
with its homotopy class. In particular, we can break up the path
into segments along which either $\s$ or $\t$ remains constant.
The composition rule for product integrals given by Eq. (11)
ensures that this break up of the Wilson loop into Wilson lines
does not depend on the intermediate points on the closed path
which are used for this purpose. So, inspired by the typical
paths which are used in the actual computations of Wilson loops
(see for
example~\cite{mal}),
we write
\beq
W = W_4\, W_3\, W_2\, W_1,
\label{w}
\eeq
In this expression, $W_k$, $k=1,..,4$, are Wilson lines such that
$\t = const.$ along $W_1$ and $W_3$, and $\s = const.$ along
$W_2$ and $W_4$.

To see the advantage of parameterizing the closed path in this
manner,
consider the  exponent of Eq. \ref{s24} :
\beq
A_a d\s^a = A_0 d\t + A_1 d\s .
\eeq
Along each segment, one or the other of the terms on the right
hand side vanishes. For example, along the segment $[\s_0 ,\s]$,
we have $\t^{'} = \t_0 = const.$.
As a result, we get for the Wilson lines $W_1$ and $W_2$,
respectively,
\beq
W_1 = \prod_{\s_0}^{\s} e^{ A_1(\sigma'; \t_0)d\sigma'}
\equiv {\cal P} e^{\int_{\s_0}^{\s}  A_1(\sigma'; \t_0)d\sigma'}
=P(\s,\s_0; \t_0),
\label{s28}
\eeq
and
\beq
W_2 = \prod_{\t_0}^{\t} e^{A_0(\s; \t')d\t'}
\equiv {\cal P} e^{ \int_{\t_0}^{\t} A_0(\s; \t')d\t'}
=Q(\s; \t,\t_0).
\label{s29}
\eeq

When the 2-surface $\Sigma$ requires more than one coordinate
patch to cover it, the connections in different coordinate
patches must be related to each other in their overlap region by
transition functions~\cite{yang}. Then, the description of Wilson
loop in terms of Wilson lines given in Eq. (26) must be suitably
augmented to take this complication into account. The product
integral representation of the Wilson line and the composition
rule
for product integrals given by Eq. (11) will still make it
possible
to describe the corresponding Wilson loop as a composite product
integral.
For definiteness, we will confine ourselves to the representation
given by Eq. (26).

It is convenient for later purposes to define two composite
Wilson line operators
$U$ and $T$ according to
\beq
U(\s, \t) = Q(\s; \t,\t_0)\, P(\s,\s_0; \t),
\label{s30}
\eeq
\beq
T(\s; \t) = P(\s,\s_0; \t)\, Q(\s_0; \t,\t_0).
\label{s31}
\eeq
Using the first of these, we have
\beq
W_2\, W_1 = U(\s, \t) .
\eeq
Similarly, we have for the two remaining Wilson lines
\beq
W_3 = P^{-1}(\s,\s_0; \t),
\eeq
and
\beq
W_4 = Q^{-1}(\s_0; \t,\t_0).
\eeq
From the Eq. (\ref{s31}), it follows that
\beq
W_4\, W_3=T^{-1}(\s,\t).
\eeq
Appealing again to Eq. (11) for the composition of product
integrals, it is clear that this expression for the Wilson loop
operator is
independent of the choice of the point ($\s,\t$).
In terms of the quantities $T$ and $U$, the Wilson loop operator
will take the compact form
\beq
W=T^{-1}(\s;\t) U(\s;\t).
\eeq

\section{Non-Abelian Stokes Theorem}
As a first step in
the proof of the non-abelian Stokes theorem, we obtain the action
of the $L$-derivative operator on $W$:
\beq
L_\t W= L_\t[T^{-1}(\s,\t)Q(\s;\t,\t_0) P(\s,\s_0;\t_0)].
\eeq
Using the
definition of the $L$-operation given by Eq. (13), noting that
$P(\s,\s_0;\t_0)$ is
independent of $\t$, and carrying out the $L$ operations on the
right hand side (RHS), we get
\begin{eqnarray}
L_\t W&=& L_\t T^{-1}(\s,\t)+T^{-1}(\s,\t)\,[\,L_\t
Q(\s;\t,\t_0)+
\nonumber\\
&&+Q(\s;\t,\t_0)
(L_\t P(\s,\s_0;\t_0))Q^{-1}(\s;\t,\t_0)]T(\s,\t).
\end{eqnarray}
Simplifying this expression by means of Eqs. (13) and (15), we
end up with
\beq
L_\t W= T^{-1}(\s,\t)[A_0(\s,\t)-L_\t T(\s,\t)]T(\s,\t).
\eeq
Next, we prove the analog of Theorem 3, which applies to an
elementary Wilson line, for the composite Wilson loop operator
defined in Eq. (24)
and made explicit in Eq. (26).

\begin{thm}
\label{thm7}

The Wilson loop operator defined in Eq. (26) can be expressed
in the form
\beq
\label{ns2}
W=\prod_{\t_0}^{\t} e^{ T^{-1}(\s, \t')[A_0(\s, \t')  - L_\t
T(\s, \t')] T(\s, \t')d\t'}.
\eeq
\end{thm}
To prove this theorem first we note from the definition of the
$L$ operation that the right hand side (RHS) of this equation can
be
written as
\beq
RHS =
\prod_{\t_0}^{\t} e^{[
T^{-1}(\s;\t')A_0(\s;\t')T(\s;\t')-T^{-1}(\s;\t'){\partial \over
{\partial \t '}}T(\s;\t')] d\t' }.
\eeq
Noting that $-T^{-1}\partial_\t T=L_\t T$, we can use
Theorem
\ref{thm5} to obtain
\beq
RHS =
T^{-1}(\s;\t)\prod_{\t_0}^{\t} e^{A_0(\s;\t')d\t'} T(\s;\t_0).
\eeq
Moreover, making use of the defining Eq. (23), we get
\beq
RHS =
T^{-1}(\s;\t)\,Q(\s;\t,\t_0)\,P(\s,\s_0;\t_0)\,Q(\s;\t_0,\t_0)
=T^{-1}(\s;\t)\,U(\s;\t).
\eeq
The last line is clearly the expression for $W$ given by Eq.
(36).

Finally, we want to
express
the Wilson loop $W$ in yet another form which we state as:

\begin{thm}
\label{thm8}

The Wilson loop operator defined in Eq. (26) can be expressed
as a surface integral of the field strength:
\beq
W=\prod_{\t_0}^{\t} e^{\int_{\s_0}^{\s} T^{-1}(\s';\t') F_{01}
(\s';\t')T(\s';\t')d\s' d\t'}
\label{s44}
\eeq
where $F_{01}$ is the 0-1 component of the non-Abelian field
strength.
\end{thm}
To prove this theorem, we note that
\beq
{\partial \over {\partial
\s}}[T^{-1}(\s,\t)A_0(\s,\t)T(\s,\t)]=
T^{-1}(\s,\t)\left[\partial_\s A_0(\s,\t)
+[A_0(\s,\t), A_1(\s,\t)]\right]T(\s,\t).
\eeq
Moreover,
\beq
{\partial \over {\partial \s}}\{T^{-1}(\s,\t)\left(L_\t
T(\s,\t)\right)T(\s,\t)\} =
T^{-1}(\s,\t)\partial_\t A_1(\s,\t)T(\s,\t).
\eeq
It then follows that
\begin{eqnarray}
&{\partial \over {\partial \s}}\{T^{-1}(\s,\t)[A_0(\s,\t)-L_\t
T(\s,\t)]\}T(\s,\t)\}
\nonumber\\
&= T^{-1}(\s,\t)[{\partial \over {\partial
\s}}A_0(\s,\t)-{\partial \over {\partial \t}}A_1(\s,\t)
+[A_0(\s,\t),
A_1(\s,\t)]\}T(\s,\t)
\nonumber\\
&= T^{-1}(\s,\t)F_{01}(\s,\t)T(\s,\t).
\label{s47}
\end{eqnarray}
The last step follows from the definition of the field strength
in terms of the connection given above
\beq
F_{0\,1}:={\partial \over {\partial
\s}}A_0(\s,\t)-{\partial \over {\partial \t}}A_1(\s,\t)
+[A_0(\s,\t),
A_1(\s,\t)].
\eeq
Integrating Eq. (\ref{s47}) with respect to $\s$, we get
\begin{eqnarray}
T^{-1}(\s,\t)[A_0(\s,\t)-L_\t T (\s,\t)]\}T(\s,\t)
\nonumber\\
= \int_{\s_0}^{\s} T^{-1}(\s';\t') F_{1\,0}
(\s';\t') T(\s';\t')d\s' d\t'.
\end{eqnarray}
We thus arrive at the surface integral representation of the
Wilson loop operator:
\beq
W=\prod_{\t_0}^{\t} e^{\int_{\s_0}^{\s} T^{-1}(\s';\t') F_{1\,0}
(\s';\t') T(\s';\t')d\s' d\t'}.
\eeq
We note that in this expression the ordering of the operators is
defined with respect to $\t$ whereas $\s$ is a parameter.
Recalling the antisymmetry of the components of the field
strength, we can rewrite this expression in terms of path ordered
exponentials familiar from the physics literature:
\beq
W={\cal P_\t} e^{ {1\over 2}\int_\Sigma
d\s^{ab}T^{-1}(\s;\t)F_{ab}
(\s;\t) T(\s;\t)},
\eeq
where $d\s^{ab}$
is the area element of the 2-surface.
Despite appearances, it must be remembered that $\s$ and $\t$
play very different roles in this expression.

\section{A Second Proof}

To illustrate the power and the flexibility of the \p\ formalism,
we give here a variant of the previous proof for the
non-Abelian Stokes
theorem. This time the proof makes essential use of the
non-trivial
Theorem 6.
We start with the form of $W$ given in Eq. (36) and take
 its derivatives with respect to $\t$:
\begin{eqnarray}
{\partial W \over {\partial \t}}&=&\partial_\t
Q^{-1}(\sigma_0;\t,\t_0)
P^{-1}(\s,\s_0;\t) Q(\s;\t,\t_0) P(\s,\s_0;\t_0)+
\nonumber\\
&&+Q^{-1}(\sigma_0;\t,\t_0)
\partial_\t P^{-1}(\s,\s_0;\t) Q(\s;\t,\t_0) P(\s,\s_0;\t_0)+
\nonumber\\
&& +
Q^{-1}(\sigma_0;\t,\t_0)P^{-1}(\s,\s_0;\t) \partial_\t
Q(\s;\t,\t_0)
P(\s,\s_0;\t_0).
\end{eqnarray}
Here, we have made use of the fact that $P(\s,\s_0;\t_0)$ is
independent
of $\t$. As a preparation for the use of Theorem \ref{thm3}, we
start with Eq. (13) for $W$ and use Theorem \ref{thm1}
\begin{eqnarray}
L_\t W={\partial W \over {\partial \t}}
W^{-1}=&T^{-1}(\s;\t)\,[A_0(\s;\t)-
P(\s,\s_0;\t)A_0(\s_0;\t)P^{-1}(\s,\s_0;\t)-
\nonumber\\
&-\partial_\t
P(\s,\s_0;\t)P^{-1}(\s,\s_0;\t)]\,T(\s;\t).
\label{n1}
\end{eqnarray}
Now we can use Theorem 6 to evaluate the derivative of
the \p\
with respect to the parameter $\t$:
\beq
\partial_\t P(\s,\s_0;\t)=\int_{\s_0}^\s
d\s'P(\s,\s';\t)\partial_\t A_1
(\s';\t)P(\s',\s_0;\t).
\eeq
Then, after some simple manipulations using the defining
equations
for the various terms in Eq. (53), we get:
\beq
T^{-1}(\s;\t)\partial_\t
P(\t)P^{-1}(\t)T(\s;\t)=\int_{\s_0}^\s d\s'
T^{-1}(\s';\t)\partial_\t A_1
(\s';\t)T(\s';\t).
\label{n2}
\eeq
Using Theorem \ref{thm2} and the fact that $P(\s_0,\s_0;\t)=1$,
we can write
the rest of
Eq. (53) as an integral too:
\begin{eqnarray}
&T^{-1}(\s;\t)[A_0(\s;\t)-
P(\s,\s_0;\t)A_0(\s_0;\t)P^{-1}(\s,\s_0;\t)]T(\s;\t)=
\nonumber\\
&=Q^{-1}(\s_0;\t,\t_0)[P^{-1}(\s,\s_0;\t)A_0(\s;\t)P(\s,\s_0;\t)
-A_0(\s_0)]Q(\s_0;\t,\t_0)=
\nonumber\\
&=\int_{\s_0}^\s d\s'\, P^{-1}(\s',\s_0;\t)(\partial_\t
A_0(\s',\t
)+[A_0(\s',\t),A1(\s',\t)])P(\s',\s_0;\t).
\end{eqnarray}
Combining Eqs. (53), (55), and (56), we obtain:
\beq
L_\t W={\partial W \over {\partial \t}}W^{-1}=\int_{\s_0}^\s d\s'
T^{-1}(\s',\t)F_{01}(\s',\t)T(\s',\t).
\eeq
Using Theorem 3, we are immediately led to Eq. (50) which was
obtained  by the previous method of proof.

There are two reasons for the relative simplicity of this proof
over the one which was given in the previous section. One is due
to the use of differentiation with respect to a parameter
according the Theorem 6. The other is due to the use of Eq. (13)
and Theorem 3 for the composite operator $W$. In the first proof,
the use of this theorem for $W$ was not assumed. Its
justification for using it in the second proof lies in the
composition law for product integrals given by Eq. (11).

\section{Convergence Issues}

The definitions of Wilson lines and Wilson loops as currently
conceived in the physics literature involve
exponentials of operators. The standard method of making sense
out of such exponential operators in the physics literature is
through their power series expansion:
\beq
{\cal P} e^{\int_a^bA(x)dx} =\sum_{n=0}^\infty {1\over
n!}\,{\cal P} \left(\int_a^bA(x)dx\right)^n,
\label{s1}
\eeq
where a typical path ordered term in the sum has the form
\beq
\frac{1}{n!}{\cal P} (\int_a^bA(x)dx)^n:=
\int_a^bdx_1\int_a^{x_1}dx_2
\ldots \int_a^{x_{n-1}}dx_n\,A(x_1)A(x_2)\ldots A(x_n).
\eeq
Without additional specifications, such a power series expansion
is purely formal, and
it is not clear {\it \'a priori} that the series
(\ref{s1}) is well defined and
convergent. Indeed, in previous attempts~\cite{rfive}-~\cite{r15}
at proving the non-abelian Stokes theorem, the convergence of
such series has been taken more or less for granted. 

One important advantage of our
product integral approach is that, without the need for further
input, we can show the convergence of
the path ordered exponentials
precise, without any further effort by using the  established 
properties of
\p s.
\footnote{For example, to estimate a general term in the
expansion (\ref{s1}) to show the norm convergence,
 we don't need to restate the elementary fact that the
space of
matrices of given rank
over the complex numbers
form a Banach space.} 
They will enable us to prove
that the series of partial sums converges uniformly
to the \p . The proof is contained, as a special case, in the
following two theorems valid for all product integrals. The
detailed proofs of these theorems are given in
reference~\cite{rfour}.

\begin{thm}
\label{thm9}
Given the continuous function $A:[a,b]\rightarrow {\bf
C}_{n\times n}$, and given
$x,y\in
[a,b]$, let $L(x,y)=\int_x^y
||A(s)||ds$. Also let
 $J_0(x,y)=I$, and for $n\geq 1 $ define iteratively
$J_n(x,y):=\int_x^y A(s) J_{n-1}(s,y)ds$. Then for any $n\geq 0$
the following holds:
\beq
||\prod_x^ye^{ A(s)ds}-\sum_{k=0}^n J_k(x,y)|| \leq {1\over
(n+1)!} |L(x,y)|^{n+1}e^{L(a,b)}.
\eeq
\end{thm}
This estimate is uniform for all $x$, $y$ in the interval
$[a,b]$. One of the consequences of this estimate is the content
of the next theorem.

\begin{thm}
\label{thm10}
With $A$ and $J_k (x,y)$ as in Theorem 9, we have, in the same
notation,

\beq
\prod_x^ye^{ A(s)ds}=\sum_{k=0}^\infty J_k(x,y).
\eeq
The series on the right hand side of this expression converges
uniformly for any
$x,y\in
[a,b]$.
\end{thm}

To give a flavor of the proofs involving \p s we will include the
proof
of this theorem. According to Theorem
\ref{thm9} we
have:
\beq
||\prod_x^ye^{ A(s)ds}-\sum_{k=0}^n J_k(x,y)||\leq M{1\over
(n+1)!} |L(b,a)|^{n+1}\longrightarrow 0,\,\, as\,
n\rightarrow \infty
\eeq
This follows immediately since, e.g., by Stirling's formula, the
asymptotic
behavior of the factorial function is roughly $n!\approx n^n$ as
$n\rightarrow \infty$. Therefore, ${x^n\over {n!}} \approx
\left({x\over n}\right)^n\rightarrow 0$ as $n\rightarrow
\infty$.
q.e.d.

\section{Physical Observables}

From the Wilson loop operator, one can obtain physical
quantities
in
a variety of ways. The most familiar one is its trace which gives
the c-number Wilson loop or the Wilson loop observable:
\beq
tr W_R(C)=tr {\cal P} e^{ {1\over 2}\int_\Sigma
d\s^{ab}T^{-1}(\s;\t)F_{ab}
(\s;\t) T(\s;\t)}.
\eeq
The subscript $R$ in this expression refers to the particular
representation of the generators.

Another invariant associated
with an operator (a matrix) is its determinant. From its product
integral representation, the determinant of the Wilson loop
operator is given by
\beq
\det W=e^{ tr{1\over 2}\int_\Sigma
d\s^{ab}T^{-1}(\s;\t)F_{ab}
(\s;\t) T(\s;\t)}.
\eeq
After some straight forward
manipulations, this can be expressed in the form
\beq
detW=e^{{1\over 2}\int_\Sigma tr
d\s^{ab}F_{ab}(\s;\t)}.
\eeq
The generators of simple Lie groups can be represented by
traceless matrices so that for these groups $tr\, F_{ab} = 0$,
indicating that $\det
W = 1$. This is not surprising since the Wilson loop operator is
a group element, and for group elements with determinant one this
result
follows trivially. For non-simple groups such as $U(1)$ and the
products thereof the trace reduces to the trace of commuting
elements of the algebra with non-zero trace. The corresponding
subgroup is commutative, there is no ordering problem, and the
surface representation of Wilson loop operator reduces to that of
the
(abelian) Stokes Theorem.

\section{The gauge transforms of Wilson lines and Wilson loops}

Under a gauge transformation, the components of the connection,
i.e. the gauge potentials, transform according to~\cite{r18}
\beq
A_\mu (x) \longrightarrow g(x)A_\mu (x)g^{-1}(x)-g(x)\partial_\mu
g(x)^{-1},
\label{gt}
\eeq
The components of the
field strength
(curvature) transform covariantly:
\beq
F_{\mu\nu}(x) \longrightarrow g(x)F_{\mu\nu}(x) g^{-1}(x).
\label{gt2}
\eeq
Using the product integral formalism, we want to derive the
effect of these gauge transformations on Wilson lines and Wilson
loops.

Let us start with the Wilson line defined by Eq. (22). Under
the gauge
transformation (\ref{gt}) this quantity transforms as
\beq
P(\sigma,\s_0;\t) = \prod_{\s_0}^{\s} e^{
A_1(\sigma';\t)d\sigma'}
\longrightarrow
\prod_{\s_0}^{\s} e^{[
g(\sigma';\t)A_1(\sigma';\t)g^{-1}(\sigma';\t)-g(\sigma';\t)
\partial_\s
g^{-1}(\sigma';\t)]d\sigma'}.
\eeq
By Eq. (13), $g(\sigma;\t)\partial_\s
g^{-1}(\sigma;\t)=-L_\s g(\sigma;\t).$
Thus, we have for the gauge transformed Wilson line
\beq
\prod_{\s_0}^{\s} e^{[
g(\sigma';\t)A_1(\sigma';\t)g^{-1}(\sigma';\t)+L_\s
g(\sigma';\t)]d\sigma'}.
\eeq
Moreover, we use Theorem 4 and recall from Theorem 3
that
$\prod_{\s_0}^{\s} e^{L_\s
g(\sigma';\t)d\sigma'}=g(\sigma;\t)g^{-1}(\sigma_0;\t).$
Then, the gauge transform of $P(\s,\s_0;\t)$ will take the form
\beq
g(\sigma;\t)g^{-1}(\sigma_0;\t)\prod_{\s_0}^{\s}
e^{g(\sigma_0;\t)A_1(\sigma';\t)g^{-1}(\sigma_0;\t)}.
\eeq
Finally, using Theorem 7 and 8 one can
readily
see that
the constant terms in the exponents can be factored out from the
\p\, so that we get
\beq
P(\sigma,\s_0;\t)\longrightarrow
g(\sigma;\t)P(\sigma,\s_0;\t)g^{-1}(\sigma_0;\t).
\label{factor}
\eeq
In the physicist's notation, the result can be stated as
\beq
{\cal P}e^{\int_a^b A_\mu (x) dx^\mu}\longrightarrow g(b)\left(
{\cal
P}e^{\int_a^b A_\mu (x) dx^\mu}\right)g(a),
\label{gt1}
\eeq
Thus, we have an unambiguous proof of how the Wilson line
transforms under gauge transformations. This is of course
consistent with the role of the Wilson line as a parallel
transport
operator. For a closed path, the points $a$ and $b$ coincide. As
a result, the corresponding Wilson loop operator transforms gauge
covariantly.

For consistency, we expect that the surface integral
representation of the Wilson loop also transforms covariantly
under gauge transformations. To show this explicitly, we note
from Eq. (50) that in this case we need to know how the operator
$T(\s,\t)$ transforms under gauge transformations. To this end,
we note that the Wilson line $Q(\sigma;\t,\t_0)$ given by Eq.
(23) transforms as
\beq
\label{factorq}
Q(\sigma;\t,\t_0) = \prod_{\t_0}^{\t} e^{
A_0(\sigma;\t')d\t'}\longrightarrow
g(\sigma;\t)Q(\sigma;\t,\t_0)g^{-1}(\sigma;\t_0).
\eeq
The transform of the composite Wilson line $T(\s,\t)$ given by
Eq. (31) follows immediately:
\beq
T(\s; \t) = P(\s,\s_0; \t)\, Q(\s_0; \t,\t_0)\longrightarrow
g(\sigma;\t)T(\s; \t) g^{-1}(\sigma_0;\t_0).
\label{gt3}
\eeq
As expected from the composition rule given by Eq. (11), the
product of two Wilson lines transforms as a
Wilson line.

From the above results, it is straight forward to show that the
surface integral representation of Wilson loop transforms as
\beq
W \longrightarrow
\prod_{\t_0}^{\t} e^{g(\sigma_0;\t_0)\left(\int_{\s_0}^{\s}
T^{-1}(\s';\t') F_{10}(\s';\t')T(\s';\t') dt'\right)
g^{-1}(\sigma_0;\t_0)}.
\eeq
As in the case of Wilson line, the constant factors in the
exponent factorize, so that under gauge transformations the
Wilson loop transforms covariantly, i.e.,
\beq
W \longrightarrow
g(\sigma_0;\t_0)\prod_{\t_0}^{\t}
e^{\int_{\s_0}^{\s} T^{-1}(\s';\t')
F_{10}
(\s';\t')T(\s';\t') dt'} g^{-1}(\sigma_0;\t_0).
\eeq
This result strengthens our confidence in the self consistency of
our formalism. In the physics notation, this
transformation law takes the form
\beq
{\cal P}e^{\oint_C A_\mu (x) dx^\mu}\longrightarrow g(a)\left(
{\cal
P}e^{\oint_C A_\mu (x) dx^\mu}\right)g^{-1}(a),
\label{gt11}
\eeq
where $a$ is a point on the loop $C$.

An important consequence of the gauge covariance of the Wilson
loop operator is that the Wilson loop observable given by Eq.
(63) is gauge
invariant. Since this observable is the trace of the Wilson loop
operator, the result follows from the invariance of trace of a
product of operators under cyclic permutation of the operators.

\section{Concluding Remarks}

The identification of Wilson lines and Wilson loops of
non-Abelian gauge
theories with product integrals allows for the possibility of 
extracting physical
consequences from these objects in a consistent and
mathematically well defined manner. Although many of the
properties of Wilson lines and Wilson loops have been
discussed~\cite{rfive}-~\cite{r15} from various, more intuitive,
points of view, there are two issues associated with these
operators with respect to which their product integral
representations have a decided advantage. One is the existence
issue discussed in Section 6, and the other is the supersymmetric
generalization of these notions~\cite{km}.
We are optimistic that the present work
will help fill the
gap in connection with these as well as other issues.

\vspace{0.3in}

This work was supported in part by the Department of Energy under
the contract number DOE-FGO2-84ER40153. The hospitality of Aspen
Center for Physics in the Summer of 1998 is also gratefully
acknowledged. We are also grateful to Dr. M. Awada for valuable
input at the initial stages of this work.

\end{document}